\documentclass[aps,preprint,preprintnumbers,superscriptaddress,
amsmath,amssymb,prl]{revtex4-2}

\usepackage[dvipdfmx]{graphicx}
\usepackage{dcolumn} 
\usepackage{bm}
\usepackage{amssymb}
\usepackage{amsmath}
\usepackage{color}

\usepackage{hyperref}

\newcommand{\FigCap}[1]{(#1)}	

\newcommand{\ts}{$T_{{\rm s}}$}
\newcommand{\tc}{$T_{{\rm c}}$}

\newcommand{\fesete}{FeSe$_{1-x}$Te$_{x}$}
\newcommand{\feses}{FeSe$_{1-x}$S$_{x}$}

\begin{document}

\title{Extended strange metal regime in a pure nematic quantum critical superconductor}

\author{Kousuke~Ishida}
\thanks{These authors have contributed equally to this work.}
\email{k.ishida@edu.k.u-tokyo.ac.jp}

\author{Kiyotaka~Mukasa}
\thanks{These authors have contributed equally to this work.}
\affiliation{Department of Advanced Materials Science, University of Tokyo, Kashiwa, Chiba 277-8561, Japan}

\author{Shusaku~Imajo}
\altaffiliation[Present address: ]{Department of Advanced Materials Science, University of Tokyo, Kashiwa, Chiba 277-8561, Japan}
\affiliation{Institute for Solid State Physics, University of Tokyo, Kashiwa, Chiba, 277-8581, Japan}

\author{Andrew~Hardy}
\affiliation{Department of Physics, University of Toronto, 60 St. George Street, Toronto, ON, M5S 1A7 Canada}

\author{Mingwei~Qiu}
\author{Mikihiko~Saito}
\affiliation{Department of Advanced Materials Science, University of Tokyo, Kashiwa, Chiba 277-8561, Japan}

\author{Aavishkar A. Patel}
\affiliation{Center for Computational Quantum Physics, Flatiron Institute, New York, New York 10010, USA}

\author{Kohei~Matsuura}
\altaffiliation[Present address: ]{Department of Applied Physics, University of Tokyo, Bunkyo-ku, Tokyo 113-8656, Japan}

\author{Yuichi~Sugimura}
\affiliation{Department of Advanced Materials Science, University of Tokyo, Kashiwa, Chiba 277-8561, Japan}

\author{Yu~Uezono}
\author{Takumi~Otsuka}
\affiliation{Graduate School of Science and Technology, Hirosaki University, Hirosaki, Aomori 036-8561, Japan}

\author{Nigel~E.~Hussey}
\affiliation{High Field Magnet Laboratory (HFML-FELIX) and Institute for Molecules and Materials, Radboud University, Toernooiveld 7, 6525 ED Nijmegen, Netherlands}
\affiliation{H. H. Wills Physics Laboratory, University of Bristol, Tyndall Avenue, Bristol BS8 1TL, United Kingdom}

\author{Takao~Watanabe}
\affiliation{Graduate School of Science and Technology, Hirosaki University, Hirosaki, Aomori 036-8561, Japan}

\author{Koichi~Kindo}
\affiliation{Institute for Solid State Physics, University of Tokyo, Kashiwa, Chiba, 277-8581, Japan}

\author{Takasada~Shibauchi}
\email{shibauchi@k.u-tokyo.ac.jp}
\affiliation{Department of Advanced Materials Science, University of Tokyo, Kashiwa, Chiba 277-8561, Japan}

\date{\today}

\begin{abstract}

High-temperature superconductivity often emerges from a strange metallic state, where the electrical resistivity exhibits a linear-in-temperature dependence over an anomalously extended temperature range. 
The prevailing belief is that magnetic critical fluctuations gives rise to strange metallicity, enhancing the superconducting transition temperature.
Here, using high pulsed magnetic fields, we have uncovered the strange metallic ground state hidden below the superconducting dome of nonmagnetic \fesete, which harbors a quantum critical point (QCP) of pure electronic nematicity, characterized by spontaneous rotational symmetry breaking.
Unlike the conventional quantum criticality, this strange metallic state does not appear in a fan-shaped region above the singular QCP but spans a wide compositional range, where pairing interactions are strengthened by nonmagnetic nematic critical fluctuations.
This stands in sharp contrast to the much cleaner system \feses, which displays a quantum critical fan above nematic QCP,
indicating that disorder-induced spatial randomness of the nematic fluctuations likely enlarges the QCP of \fesete\, into an extended region of criticality, as suggested by relevant hybrid Quantum Monte Carlo simulations.
These observations highlight superconductivity promoted by a unique interplay between pure nematic critical fluctuations, strange metallicity and disorder, providing new insight into the emergence of non-Fermi-liquid transport in various correlated materials.

\end{abstract}

\maketitle

\clearpage

\section{Introduction\label{sec:intro}}
Within Landau's Fermi liquid framework, electron-electron scattering leads to electrical resistivity with a quadratic temperature dependence, as demonstrated in transitional metals \cite{Rice1968} and certain correlated electron systems \cite{Kadowaki1986}.
This is a fundamental consequence of the Pauli exclusion principle, which constrains  the electrons participating in the electron-electron collisions to those confined within a thermal window  near the Fermi level.
In contrast, a wide variety of unconventional superconductors, including copper oxides \cite{Cooper2009,Jin2011}, iron pnictides \cite{Shibauchi2014}, heavy fermions \cite{Custers2003}, and organic Bechgaard salts \cite{Nicolas2009}, display resistivity with a linear temperature dependence down to the lowest temperatures studied. 
This unusual normal state, referred to as strange metal, often emerges above the superconducting transition temperature \tc, pointing to their intimate connection \cite{Phillips2022}.
Nevertheless, the linear temperature dependence of electrical resistivity cannot be straightforwardly captured by the picture of scattering between quasiparticles.
This puzzle has sparked extensive experimental and theoretical studies to uncover its microscopic origin.

The strange metal has often been discussed in relation to the quantum critical point (QCP), where a continuous transition is tuned to absolute zero temperature by a non-thermal external parameter.
In the phase diagram of quantum critical systems, alongside the thermal length, there exists a dynamical correlation length associated with the non-thermal external parameter, which diverges toward the QCP.
The coexistence of these two length scales results in a funnel-shaped region above the QCP, wherein quantum critical fluctuations have a substantial influence throughout the entire system.
In iron pnictides \cite{Shibauchi2014} and heavy fermions \cite{Custers2003} strange metallic behavior is observed inside a so-called quantum-critical fan above a QCP associated with antiferromagnetic order.
Away from the QCP, Fermi-liquid-like $T^2$ resistivity is restored at low temperatures.
Upon approaching the critical point, the onset temperature of $T^2$ resistivity decreases and the coefficient of the $T^2$ term, a measure of the electron correlation, diverges as a consequence of quasiparticles coupled to the enhanced critical fluctuations of the order parameter.
While these features are documented as characteristic of the resistivity linked to the classic quantum criticality, it has been increasingly recognized that in some systems such as overdoped cuprates \cite{Cooper2009,Jin2011} and organic superconductors \cite{Nicolas2009}, strange metallic transport down to low $T$ is observed over an extended range of the phase diagram rather than at a singular point above the QCP.
The coefficient of $T$-linear resistivity in the broad strange metal regime appears to scale with \tc, implying their direct link \cite{Cooper2009,Nicolas2009,Yuan2022}.

Understanding the strange metals that are seen in these quantum materials  has been a major theoretical challenge in correlated electron physics.
Recent theoretical works shows that disorder, present within all the real materials, plays a crucial role in strange metallicity.  
Including disorder-induced spatially inhomogenous interactions in models of quantum criticality produces strange metal transport that is unobtainable in translationally-invariant theories due to constraints on momentum relaxation \cite{Patel2023,Patel2024,Patel2025}.
Furthermore, strong disorder effects have been shown to replace the singular QCP in these models with an extended strange metal regime \cite{Patel2024,Patel2025,Radaelli2025}. 
However, these theoretical discussions have been largely limited to those systems whose phase diagram includes long-range magnetic order, leading to the long-standing conjecture that spin fluctuations are their essential origin.
Therefore, it has not yet been fully experimentally established whether fluctuations of nonmagnetic order can give rise to strange metallic behavior.

The iron chalcogenide FeSe undergoes an electronic nematic phase transition at \ts\, $\sim$ 90\,K, which breaks the four-fold rotational symmetry of the lattice \cite{Shibauchi2020}.
In contrast to other nematic orders, electronic nematicity of FeSe does not intertwine with magnetic or any other long-range orders.
Upon Te substitution, the nematic phase transition of \fesete\, is continuously suppressed and eventually vanishes at a nonmagnetic nematic QCP with $x_{\rm c}\sim0.50$, where the nematic susceptibility displays a strong Curie-Weiss type divergence \cite{Mukasa2021,Ishida2022}.
Remarkably, the superconducting dome of \fesete\, straddles the underlying nematic QCP even when the dome contracts under the high magnetic field, suggesting that the pairing interaction is strengthened by nematic critical fluctuations \cite{Mukasa2023}.
These nonmagnetic critical fluctuations may also influence low-energy excitations of the quasiparticles around the Fermi surface in the normal state, potentially giving rise to the marked deviation from Fermi liquid-like metallic behavior.
Here, we explore this issue by exposing the low-temperature normal state resistivity masked by the superconducting transition and we present a comprehensive investigation into the low-temperature resistivity of non-magnetic \fesete\, superconductors across various compositions, where the superconductivity is fully suppressed by applying pulsed magnetic fields up to 60\,T.

\begin{figure}[t]
\centering
\includegraphics[width=\linewidth]{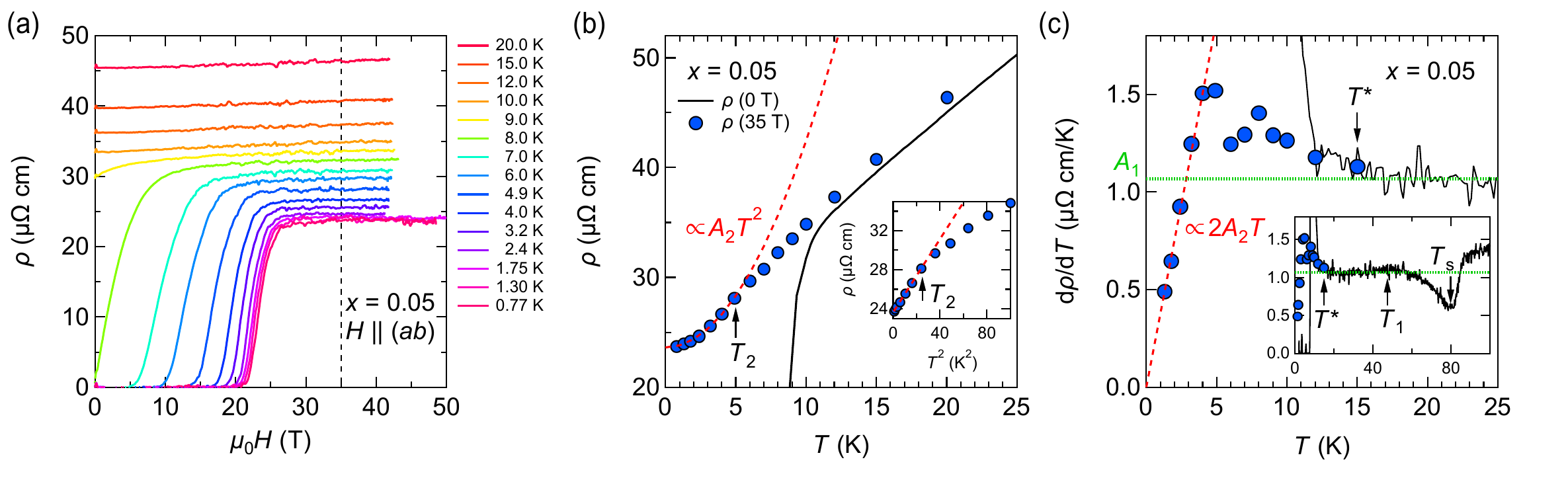}
\caption{\label{fig:MR}
Longitudinal high-field magnetoresistance and low-temperature $T^2$-resistivity masked by superconductivity in \fesete\ with $x=0.05$.
\FigCap{a} Longitudinal magnetoresistance of \fesete\, with $x=0.05$, in which the current and field are both applied within the $ab$-plane.
\FigCap{b} Temperature dependence of the resistance under zero field ($\rho (0 \rm{T})$, black curves) and at 35\,T ($\rho (35 \rm{T})$, closed blue circles) interpolated from the dataset plotted in Fig.\,\ref{fig:MR}\FigCap{a} (vertical black dashed line). 
The inset shows the low-$T$ data plotted against $T^2$ to determine the temperature scale $T_{2}$, below which resistivity follows a quadratic temperature dependence.
The red dashed line represents the obtained fitting function.
\FigCap{c} Corresponding temperature derivatives of $\rho (0 \rm{T})$, $\rho (35 \rm{T})$, and the $T$-square fitting curves shown in Fig.\,\ref{fig:MR}\FigCap{b}.
The same data in a wider temperature window are shown in the inset.
Green horizontal dashed line highlights the $T$-linear resistivity regime, in which $d\rho/dT$ has an approximately constant value $A_{1}$.
Important temperature scales are indicated by black arrows: the temperature $T^*$($T_{1}$), above(below) which $\rho(T)$ becomes $T$-linear and nematic transition temperature $T_{\rm s}$.
}
\end{figure}

\section{Results}

Due to its semimetallic electronic structure, the transverse magnetoresistance ($I\parallel (ab)$, $H\parallel c$) of \fesete\, is expected to be sizable and certainly much larger than the corresponding longitudinal magnetoresistance ($I\parallel (ab)$, $H\parallel (ab)$).
To discuss the intrinsic metallic resistivity masked by superconductivity below \tc\, from the high-field data, we focus here on a systematic study of the longitudinal magnetoresistance, which is found to be negligible or small across the entire composition range covered in the present study.
Figure\,\ref{fig:MR}\FigCap{a} shows a representative dataset of the longitudinal magnetoresistance for $x=0.05$.
The resistivity values at 35\,T obtained directly from these curves appear to be vertically shifted from the zero-field resistivity data, as depicted in Fig.\,\ref{fig:MR}\FigCap{b}.
Below $T_{2} \sim 5\,{\rm K}$, the resistivity at 35\,T is linear when plotted versus $T^2$, implying that in this regime the resistivity has a quadratic temperature dependence ($\rho(T)=A_{0}+A_{2}T^{2}$).
This becomes more apparent upon inspection of the temperature derivative $d\rho/dT$ shown in Fig.\,\ref{fig:MR}\FigCap{c}, which is $T$-linear below $T_{2}$ with a zero intercept ($d\rho/dT=2A_{2}T$).
Thus, the electronic ground state for $x=0.05$ is consistent with a correlated Fermi liquid description.
At higher temperatures above $T^* \sim 15\,{\rm K}$, $d\rho/dT$ is approximately constant up to $T_{1} \sim 47\,{\rm K}$, demonstrating the crossover from $T^2$ to $T$-linear resistivity.
We note that similar $T^2$ resistivity masked by superconductivity with comparable $T_{2}$ values and its crossover to $T$-linear resistivity has previously been reported in FeSe \cite{Licciardello2019}. 

\begin{figure}[t]
	\centering
	\includegraphics[width=\linewidth]{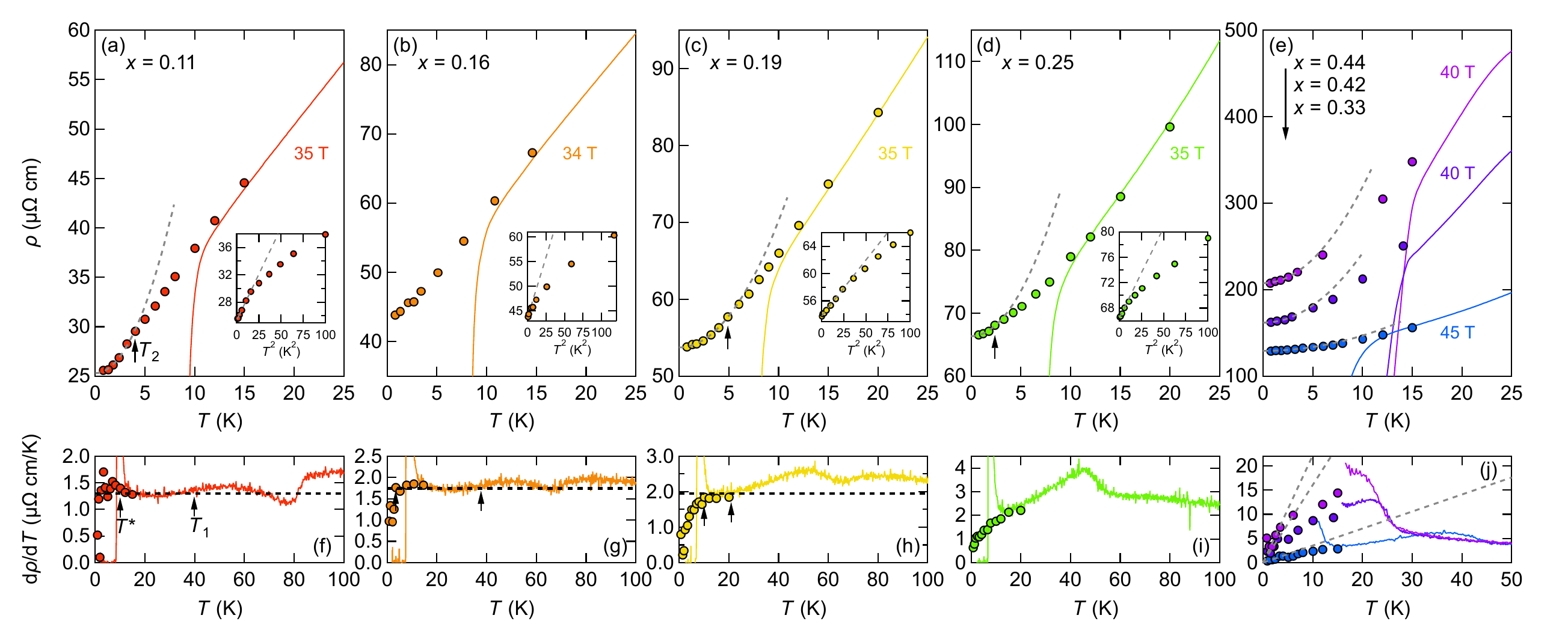}
	\caption{\label{fig:rho_nematic}
	Electrical resistivity inside the nematic phase of \fesete.
	The upper panels depict the zero resistivity curves with high field data (closed circles) for \FigCap{a} $x=0.11$, \FigCap{b} $x=0.16$, \FigCap{c} $x=0.19$, \FigCap{d} $x=0.25$, and \FigCap{e} $x=0.33, 0.42, 0.44$.
	The insets of \FigCap{a}-\FigCap{d} show the plot against $T^2$ used for determining $T_{2}$ denoted by black arrows.
	Gray dashed lines represents the fitting function of the $T^2$ resistivity.
	Each lower panel \FigCap{f}-\FigCap{j} displays the corresponding temperature derivatives $d\rho/dT$.
	For \FigCap{f}-\FigCap{h}, the $T$-linear resistivity regime is marked by horizontal dashed lines and two black arrows which represent $T^{*}$ and $T_{1}$.
	To illustrate the enhancement of $A_{2}$ with increasing $x$, the lines of $2A_{2}T$, corresponding to the temperature derivatives of the $T^2$ fitting functions shown in \FigCap{e}, are displayed in \FigCap{j}.
	}
	\end{figure}

\begin{figure}[t]
	\centering
	\includegraphics[width=\linewidth]{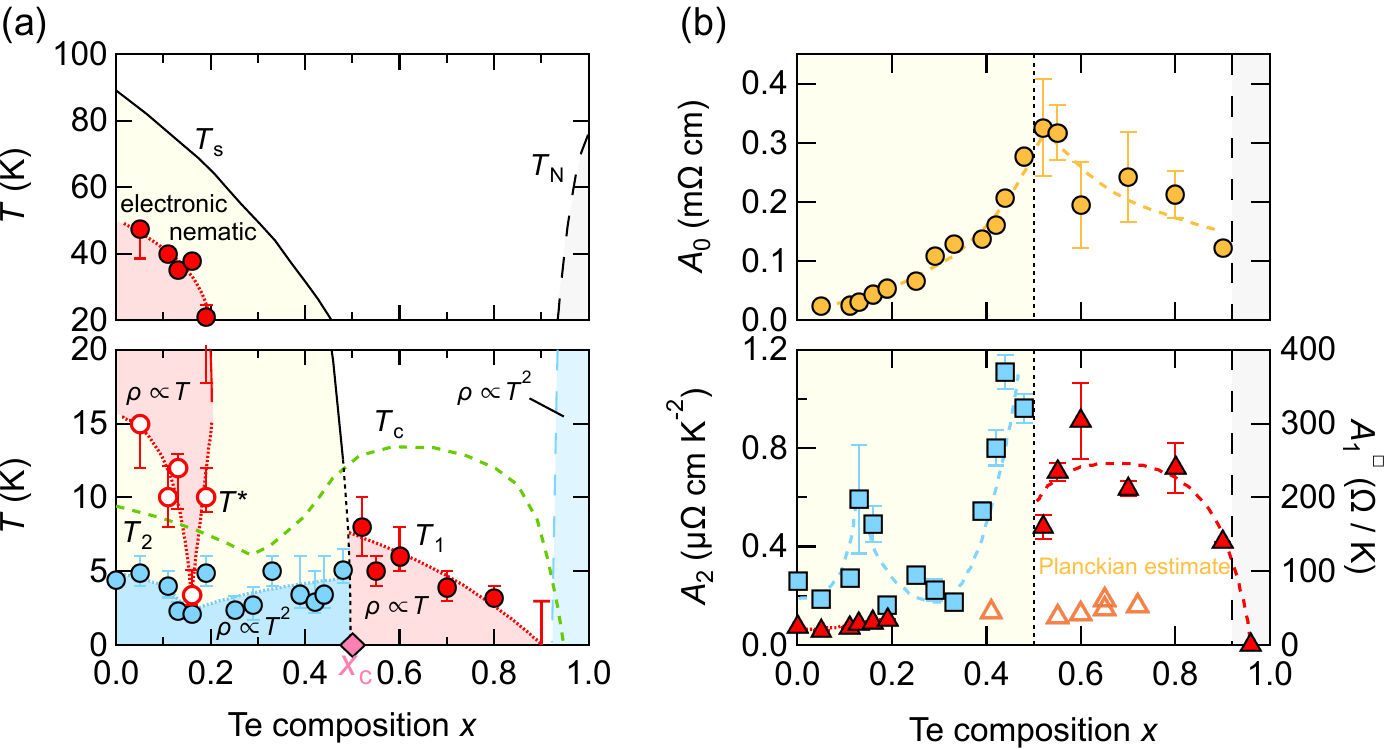}
	\caption{\label{fig:PD}
	Strange metal regime hidden inside the superconducting dome of \fesete.
	\FigCap{a} Te compositions $x$ versus temperature phase diagram at low (bottom panel) and high temperatures (top panel).
	Closed blue and red circles denote temperature $T_{2}$ and $T_{1}$ below which the resistivity $\rho$ follow $T^2$ and $T$-linear dependence, respectively ($\rho \propto T^2$ and $\rho \propto T$).
	Open red circles represent the crossover temperature $T^{*}$ between these two regimes.
	$T_{\rm s}$ and $T_{\rm N}$ represent the nematic and antiferromagnetic transition temperatures, while $x_{\rm c}$ corresponds to the nematic QCP.
	Zero-field superconducting transition temperature \tc\,is schematically depicted by green dashed curve. 
	\FigCap{b} $x$ dependence of $A_{2}$ (blue squares), $A_{1}^{\Box}$ (red triangles), and $A_{0}$ (yellow circles).
	The experimental values of $A_{1}^{\Box}$ are compared with the Planckian estimates (open orange triangles), calculated as $A_{1}^{\Box} = (h/2e^2)(1/\sum_{i} T_{\rm F}^{i})$.
	All the lines are guides to the eye.
	}
	\end{figure}

Following similar procedures of the magnetoresistance measurements and analysis, we have studied the evolution of low-temperature resistivity of \fesete\, with Te substitution spanning from $x=0.05$ to $x=0.90$ (See Supplementary Information for all magnetoresistance data and subsequent analysis).
Figure\,\ref{fig:rho_nematic} summarizes the essential features observed inside the nematic phase ($x<0.50$).
While resistivity in all compositions follows a $T^2$ dependence at low temperatures (it is linear when plotted as a function of $T^2$ and equivalently $d\rho/dT$ is linear with zero intercept), their temperature derivatives, shown in the bottom panels of Fig.\,\ref{fig:rho_nematic}, reveal that the crossover to a pure $T$-linear behavior at $T^*$, manifest as a constant $d\rho/dT$, disappears for $x\ge0.25$.
Around $x=0.16$, $T_{2}$ and $T^*$ are most strongly suppressed, resulting in a $T$-linear resistivity regime that extends down to the lowest temperature.
Upon approaching the QCP with $x_{\rm c}\sim0.50$, the $T^2$ behavior becomes more pronounced with progressive enhancement of $A_{2}$, as shown in the bottom panel of Fig.\,\ref{fig:PD}\FigCap{b}.
This is highlighted in Fig.\,\ref{fig:rho_nematic}\FigCap{j}, which shows that the slope of $d\rho/dT$ at low temperature gets steeper as $x \rightarrow x_{\rm c}$.
Figure\,\ref{fig:rho_nematic} also shows that the residual resistivity $A_0$ raises rapidly as $x$ increases, a point we shall return to later.

\begin{figure}[t]
	\centering
	\includegraphics[width=\linewidth]{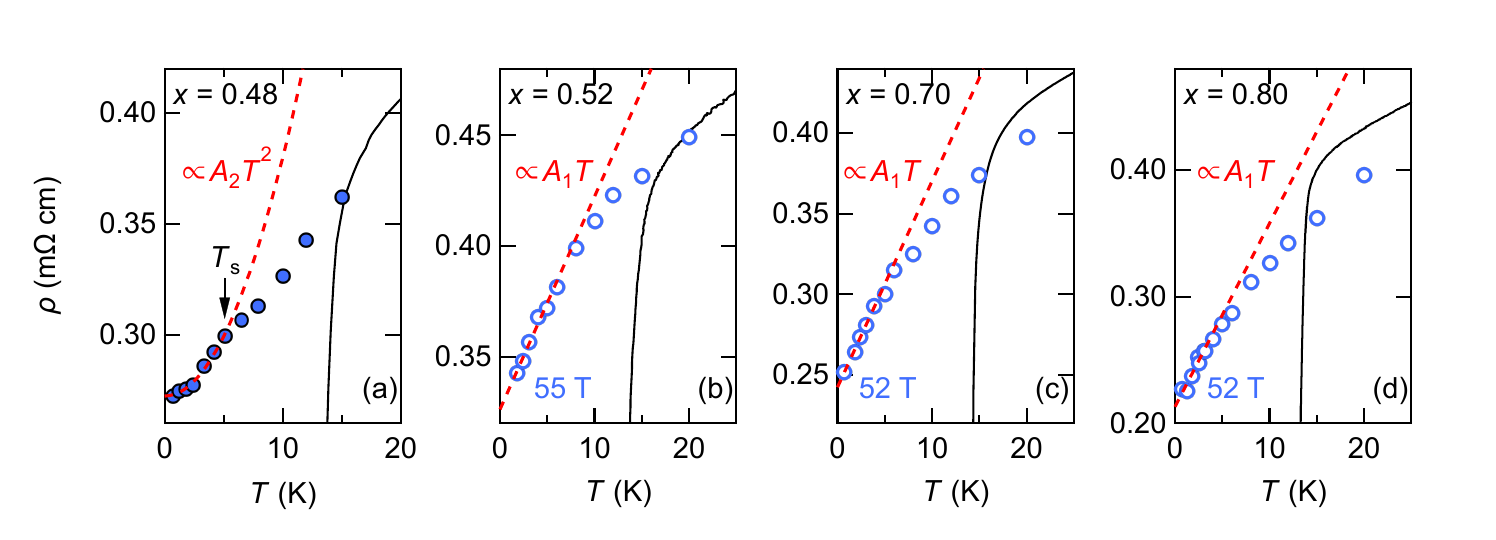}
	\caption{\label{fig:rho_QCP}
	Electrical resistivity across the nematic quantum critical point of \fesete.
	Resistivity curves under zero field (black curves) and resistivity data below \tc, uncovered by high-field magnetoresistance measurements for \FigCap{a} $x=0.48$, \FigCap{b} $x=0.52$, \FigCap{c} $x=0.70$, and \FigCap{d} $x=0.80$. 
	For $x=0.48$, resistivity values below \tc, are linearly extrapolated from high-field data (closed blue circles). For others, the high-field resistivity values directly interpolated from magnetoresistance data are shown as open blue circles. 
	The red dashed curve and lines represent fitting functions that best describe the low-temperature resistivity.
	}
	\end{figure}

Figure\,\ref{fig:rho_QCP} presents the resistivity across the nematic QCP with $x_{\rm c}\sim0.50$.
For $x=0.48$, which is located just below $x_{\rm c}$, a signature of the nematic transition could only be discerned by suppressing \tc\, in the high magnetic field.
The resistivity data below 5\,K again displays $T^{2}$ dependence albeit with an enhanced $A_{2}$ value, indicative of a correlated Fermi liquid ground state.
Beyond nematic QCP ($x>0.50$), the form of the low-$T$ resistivity changes significantly.
The resistivity curve for $x=0.52$ appears to exhibit a linear temperature dependence ($\rho(T)=A_{0}+A_{1}T$) below $T_{1} \sim 8\,{\rm K}$, in stark contrast to what is observed in $x=0.48$.
Above $T_{1}$, it deviates from the $T$-linear form and asymptotically approaches a saturation value that coincides with the Mott-Ioffe-Regel limit.
Notably, when moving away from the nematic QCP, the low-temperature resistivity does not recover a $T^2$ dependence characteristic of a Fermi liquid ground state, at least down to the base temperature of our measurements.
Instead, the $T$-linear form of the resistivity is found to persist up to $x=0.90$ (See Supplementary Information for plots of the low-temperature, high-field resistivity data for other $x$).
We note that $T_{1}$ gradually decreases with increasing $x$, being suppressed below 3\,K at $x=0.90$.

\section{Discussion\label{sec:discussion}}
Our key finding is the persistence of $T$-linear resistivity in \fesete\ extending beyond the pure nematic QCP across a broad range of compositions.
The onset temperature $T_{1}$ is far below the Debye temperature $\Theta_{\rm D} \sim$ 140\,K \cite{Klein2010}, above which the electron-phonon scattering is expected to yield a $T$-linear resistivity, confirming the non-phononic origin of this behavior.
We thus presume here that the observed low-temperature $T$-linear resistivity most likely arises from electron-electron scattering dressed by bosonic fluctuations, which soften to a sufficiently low energy relative to the given temperature scale.
Although the $T$-linear resistivity regime is sandwiched between the nematic and magnetic phases, several existing results support the notion that nematic fluctuations are playing the dominant role.
The magnetic phase of FeTe features a bicollinear spin structure distinct from that present in the iron pnictide superconductors, with sharp jumps in several physical quantities across the phase transition, indicative of a first-order transition \cite{Li2009}. 
Crucially, a nuclear magnetic resonance study within the $T$-linear resistivity regime has shown the absence of sizable low-energy spin fluctuations that could affect the charge transport probed in this study \cite{Arcon2010}.
In contrast, the nematic phase spans a broad range of the phase diagram, as illustrated in Fig.\,\ref{fig:PD}\FigCap{a}.
The putative nematic transition hidden below the superconductivity at $x=0.48$ unmasked by the present measurements (Fig.\,\ref{fig:rho_QCP}\FigCap{a}) is consistent with a QCP being located at $x_{\rm c}\sim0.50$.  
Elastoresistivity studies have demonstrated that the nematic susceptibility of the  corresponding symmetry channel follows a Curie-Weiss law above \ts\, \cite{Ishida2022}.
This behavior persists outside of the nematic phase, where $T$-linear resistivity is observed at low temperatures \cite{Jiang2023}.
Upon approach to the nematic QCP from the disordered side, nematic fluctuations are enhanced in close correspondence with the gradual increase of $T_{1}$ \cite{Jiang2023}.
Furthermore, as demonstrated by the field-insensitive critical softening of the elastic constant associated with the nematic susceptibility \cite{Goto2011}, magnetic fields in nonmagnetic \fesete\, do not significantly affect the underlying nematic quantum criticality.
This supports a direct link between the emergence of $T$-linear resistivity under high fields shown in Fig.\,\ref{fig:rho_QCP} and nematic critical fluctuations.
These combined factors indicate that the present observation can be taken as evidence for an extended strange metal regime driven by quantum nematic fluctuations.

There also exists a crossover from $T^2$ to $T$-linear resistivity deep inside the nematic phase, the underlying origin of which remains elusive.
Putative strong short-range fluctuations might be linked to this crossover, but the relevant order parameter has yet to be identified.
It is unlikely that nematic fluctuations could be responsible here as they should be suppressed sufficiently below \ts.
Although at present it remains unclear how the low-energy spin fluctuations of \fesete\, evolve in this region, pressure-induced magnetic order rapidly shrinks with increasing Te compositions \cite{Mukasa2021}, implying that magnetic correlations at ambient pressure become weaker.
The monotonic reduction of \tc\, in \fesete\, up to $x\sim0.30$ also supports the suppression of spin fluctuations, which likely plays a dominant role in electron pairing in this low-$x$ region \cite{Wiecki2018,Mukasa2023}.
The anomalous transport behavior found inside the nematic phase warrants further experimental and theoretical investigation.

In Fig.\,\ref{fig:PD}\FigCap{b}, we show the $x$ dependence of $A_{1}^{\Box}\equiv A_{1}/d$, where the coefficient $A_{1}$ is normalized by the separation between conducting planes.
It reveals that the scattering rate in the $T$-linear resistivity regime approximately scales with the \tc\, dome, suggesting that strange metallicity is intimately related to superconductivity in these compositions.
This is in line with the fact that in non-superconducting bulk \fesete\, single crystals, low-temperature resistivity recovers a quadratic temperature dependence inside the antiferromagnetic phase, consistent with a Fermi liquid ground state \cite{Otsuka2019,Sato2025} (See Supplementary Information for a discussion of the $T^2$ resistivity in highly substituted \fesete.).
In this context, the recent report of superconductivity in FeTe thin films \cite{Yan2026} motivates investigation of the strange metallicity below \tc\, in these films, although it remains unclear whether such superconductivity can be realized in bulk single crystals.
It is striking here that $A_{1}^{\Box}$ of optimally substituted \fesete\, is an order of magnitude larger than the values in FeSe, whose scattering rate is bounded by Planckian dissipation as $\hbar/\tau = \alpha k_{\rm B}T$ with $\alpha \sim 1$ \cite{Licciardello2019}.
So far, several correlated systems exhibiting $T$-linear resistivity appear to have $\alpha \sim 1$, suggesting that this limit might be a generic property, irrespective of the microscopic scattering mechanism \cite{Bruin2013}.
Such a steep linear slope in \fesete\, indicates that either the scattering rate within the extended strange metal regime notably can exceed this Planckian limit by a considerable amount ($\alpha \gg1$) or that the effective Plasma frequency $\Omega_{\rm p}$ in \fesete\, is markedly reduced.
Indeed, in quasi-two-dimensional metals with Planckian dissipation, $A_{1}^{\Box}$ scales as $A_{1}^{\Box}\propto 1/T_{\rm F}\propto m^{*}/n$, where $T_{\rm F}$, $m^{*}$, and $n$ represent Fermi temperature, effective mass and carrier density, respectively \cite{legros2019}.
While angle-resolved photoemission spectroscopy (ARPES) measurements have revealed that the effective mass of the $d_{xy}$ band around $\Gamma$ point for $x=0.50-0.60$ is about twice as large as that for FeSe \cite{Huang2022}, there has been no corroborating evidence for a sufficiently large decrease in the carrier density, which is required for the scattering rate to remain at Planckian limit.
To clarify this point, we calculated $A_{1}^{\Box}$ under the assumption that the scattering rate is at the Planckian bound ($\alpha =1$) as $A_{1}^{\Box} = (h/2e^2)(1/\sum_{i} T_{\rm F}^{i})$ with $T_{\rm F}$ values extracted from ARPES studies \cite{Lubashevsky2012,Rinott2017,Liu2015} (See Supplementary Information for more details of this procedure).
The Planckian estimates are of the same order as the observed $A_{1}^{\Box}$ in the low-$x$ region but fall significantly below the values obtained beyond $x_{\rm c}$ (bottom panel of Fig.\ref{fig:PD}\FigCap{b}).
The mechanism of this super-Planckian transport (with $\alpha \gg1$) beyond pure nematic QCP remains an open question, but suggests that once the charge dynamics become quantum critical, $\alpha$ is bound by instabilities that lead to phase transitions rather than a fundamental bound based on universal constants \cite{Murthy2023, Hardy2025}.

While the extended $T$-linear resistivity regime conflicts with the conventional picture of quantum criticality, upon approaching the nematic QCP from the ordered state, the coefficient $A_{2}$ shows a sharp five-fold increase, as shown in the bottom panel of Fig.\,\ref{fig:PD}\FigCap{b}.
A recent ARPES study showed that upon approaching the QCP with increasing $x$, the flat $d_{xy}$ hole band becomes more correlated and shifts towards the Fermi level.
It should be noted, however, that this hole band is still located 20\,meV below  Fermi level at $x=0.40$ \cite{Morfoot2023}.
Given that the resistivity follows a $T^2$ dependence down to the lowest temperature $\sim700$\, mK, it is unlikely that this increase of $A_{2}$ originates solely from an increase in the correlation strength in the $d_{xy}$ band.
Rather, it may reflect the enhancement of electron-electron scattering by singular nematic critical fluctuations, though, the slight stabilization of the onset temperature $T_{2}$ as $x \rightarrow x_{\rm c}$ appears incompatible with a conventional quantum critical scenario in which this temperature scale vanishes.

In contrast to the anomalously extended strange metal regime found in \fesete, the nematic quantum criticality of \feses\, appears to fit a more conventional picture, in which $T$-linear resistivity exists only within the funnel-shaped region above the nematic QCP \cite{Licciardello2019}.
This raises an important question on the mechanism underlying the contrasting transport signatures of nematic quantum criticality in these systems.
One potentially relevant distinction between \feses\, and \fesete\, is that Curie-Weiss temperature of \fesete\, derived from nematic susceptibility remains approximately zero over a wide compositional range beyond $x_{\rm c}$ \cite{Jiang2023}, whereas in \feses, it decreases monotonically once the nematic QCP is crossed \cite{Hosoi2016,Ishida2022}.
Another notable difference is the effect of the isovalent substitutional disorder, as discussed in detail below.
In \feses, substitution with the lighter element produces a comparatively pristine system, reflected in the ability to observe quantum oscillations across the nematic QCP \cite{Coldea2019} and a residual resistivity that is around one order of magnitude lower than found in \fesete\, at equivalent $x$ values (top panel of Fig.\,\ref{fig:PD}\FigCap{b}).
In \fesete, the nematic QCP at $x_{\rm c} \sim 0.50$ coincides with the highest level of isovalent substitutional disorder.
Moreover, laser-excited photoemission electron microscopy measurements \cite{Shimojima2021,Kageyama2024} directly show that the isovalent substitution in FeSe reduces electronic nematic domain size, demonstrating that this system includes the spatial disorder in the electronic nematic order parameter itself. 
Motivated by this, we consider a model of fermions coupled to the substitution-induced spatially random electronic nematic interactions (See Supplementary Information).
The effective theory shows that spatial disorder in the order parameter causes the associated fluctuations couple to the entire Fermi surface, washing out particular details of the ordering wavevector or the electronic structure of the materials \cite{Patel2024}.
Exact numerical results of hybrid Monte Carlo simulation for a similar model where fermions couple to antiferromagnetic fluctuations \cite{Patel2025} show that spatial disorder can cause an extended Griffiths phase to develop around or instead of a quantum critical point.
Taken together, these results suggest that incorporating the effects of isovalent disorder in the electronic nematic interactions can also stabilize such an extended Griffiths phase.
Such extended Griffiths phases have been shown to provide a mechanism for extended regions of strange metal behavior in related theoretical works \cite{Patel2024,Patel2025}.

Substitution-induced potential disorder also provides a qualitative explanation of the observed sharp peak in the residual resistivity $A_0$ around the pure nematic QCP, shown in the top panel of Fig.\,\ref{fig:PD}\FigCap{b}.
By computing a toy model of fermions with substitutional disorder, we find that peak structure of $A_{0}$ itself can be explained by single-particle disorder physics (See Supplementary Information), although the experimentally observed sharpness and dramatic increase near $x_{\rm c}$ hint at additional mechanisms.
It should be noted here that the peak in $A_{0}$ cannot be attributed to the divergence of the effective mass $m^*$ due to electron interactions, inferred from the increase of $A_{2}$, because the non-interacting mass $m_{e}$ is what appears in expressions for conductivity due to impurity scattering. 
Furthermore, perturbative calculations show that Altshuler–Aronov corrections from quantum critical nematic fluctuations reduce rather than enhance the residual resistivity \cite{Hartnoll2014}.
One possible mechanism of increase in $A_{0}$ beyond the single-particle disorder effect is the emergence of glassy puddles of local nematic order formed by spatially random nematic interactions, which act as sources of quasi-elastic scattering.
In addition to their effect on residual resistivity, such spatially random disorder would play an important role in relaxing electron momentum over the nematic quantum critical regime.
Since nematic fluctuations associated with a $\bm{Q}=0$ order parameter involve only small momentum transfer, Umklapp scattering processes are expected to be negligible in \fesete\, with relatively small Fermi surfaces.
Spatially random nematic critical interactions, however, would acquire a broad wavevector distribution that couples to multiple electron momenta, thereby providing a source of momentum relaxation strong enough to generate $T$-linear resistivity.
This mechanism yields persistent scattering between the critical fluctuations and the electronic degree of freedom down to zero temperature \cite{Patel2023,Patel2025}, and is conceptually distinct from a semiclassical Boltzmann-equation approach in the presence of impurity scattering, which predicts a $T^{4/3}$ dependence \cite{Dell'Anna2007,Dell'Anna2007e,Carvalho2019}.
These considerations call for the more systematic study of models of nematic quantum criticality with substitution-induced interaction and potential disorder.

The key premise of the present observation is that nonmagnetic nematic fluctuations can lead to an extended strange metal regime, which correlates with electron pairing.
This observation may have important implications for our understanding of the anomalous low-temperature $T$-linear resistivity seen in other correlated systems, especially the long-debated strange metal regime in cuprates.
In overdoped hole-doped cuprates, linearity of the low-temperature resistivity persists to the edge of the superconducting dome with a coefficient $A_{1}$ scaling with \tc, and beyond the doping where the superconducting dome terminates, the temperature dependence of the resistivity becomes purely $T^2$ \cite{Cooper2009}.
The extended $T$-linear resistivity regime emerges beyond the endpoint of the pseudogap temperature line intersecting the superconducting dome, which does not coincide with the critical doping of magnetic order.
These striking similarities to the phase diagram of \fesete\, with nonmagnetic QCP clearly motivate further investigation.

\section*{Methods}
\subsection*{Single crystals}
Single crystals of \fesete, with $0 < x \leq 0.48$ were synthesized by the chemical vapor transport (CVT) technique \cite{Mukasa2021}, while those with $0.52 \leq x \leq 0.90$ were prepared using the Bridgman method \cite{Watanabe2020}.
To suppress the effect of excess Fe, crystals grown via the Bridgman method underwent a Te annealing process \cite{Watanabe2020}.
Previous studies \cite{Koshika2013,Roppongi2025} have shown that, with sufficient Te annealing, the amount of excess Fe becomes negligible: the resistivity exhibits fully metallic behavior in contrast to the insulating behavior of as-grown crystals, and the diamagnetic response becomes sharp with nearly 100\,\% superconducting volume fraction.
Consistent with these established results, our Te-annealed crystals display metallic transport down to the lowest measured temperatures, without any low-temperature upturn in the resistivity. 
This behavior indicates that excess Fe has been effectively removed in our samples.
We also note here that a recent study \cite{Huang2022AIP} shows that single crystals with $x = 0.55$ synthesized by the CVT method exhibit properties comparable to those of Bridgman-grown crystals subjected to a post annealing procedure.
Furthermore, the absence of any obvious enhancement of the residual resistivity $A_{0}$ beyond $x_{\rm c}$ indicates that Bridgman-grown crystals are not significantly more disordered than CVT-grown ones. 
Thus, differences in growth method are unlikely to significantly affect the transport properties reported in this work.

For samples obtained by CVT, the actual Te concentration $x$ was determined for each crystal from the $c$-axis lattice parameter measured by X-ray diffraction.
The Bridgman-grown crystals were characterized using the nominal Te composition values.
    
\subsection*{High-field resistivity measurements}
Resistivity measurements under high magnetic fields were carried out at the International MegaGauss Science Laboratory, Institute for Solid State Physics, University of Tokyo, using a pulsed magnet with a maximum field of 60 T.
The electrical resistivity were measured using the standard four-probe configuration, with current applied within the $ab$-plane.
In all the measurements, the magnetic field was applied parallel to the $ab$-plane.

\section*{Data Availability}
The data that support the findings of this study are available from the corresponding authors upon reasonable request.

\section{acknowledgments}
We thank H. Kontani for the fruitful discussion.
This work was partially carried out by the joint research in the Institute for Solid State Physics, University of Tokyo. 
This work was supported by Grants-in-Aid for Scientific Research (KAKENHI Grant Nos.\,JP22H00105 and JP23H00089) and Grant-in-Aid for  Transformative Research Areas (A) “Correlation Design Science” (KAKENHI Grant No. JP25H01248) from Japan Society for the Promotion of Science, and the Engineering and Physical Sciences Research Council (EPSRC Grant No. EP/V02986X/1).

\bibliographystyle{apsrev4-2}
\bibliography{FeSeTe_transport.bib}

\end{document}